# Towards a Monotonicity-Compliant Price Index for the Art Market

Ventura CHARLIN [1] and Arturo CIFUENTES [2]

*[1] V.C. Consultants, Santiago, CHILE; ventcusa@gmail.com*

*[2] University of Chile, Santiago, CHILE; arturo.cifuentes@fen.uchile.cl*

## Working Paper March 18, 2014

**SUMMARY**

Notwithstanding almost forty years of efforts, the market for paintings still lacks a widely accepted price index. In this paper, we introduce a simple and intuitive metric to construct such index. Our metric is based on the price of a painting divided by its area. This formulation rests on a solid mathematical foundation as it corresponds to a particular type of hedonic model. However, unlike indexes based on the time-dummy coefficients of conventional hedonic models, this index satisfies the monotonicity condition. We demonstrate with a simple example the advantages of our metric. We also show the dangers of relying on the time-dummy coefficients of conventional hedonic models to estimate returns and generate price indexes.

# INTRODUCTION

All mature markets are alike: they have at least one (and often many) widely accepted and transparent index that tracks market movements. Frequently, such indexes are used to build derivative contracts. Clearly, the art market —and more specifically, the market for paintings— is still in its infancy. Despite almost forty years of both academic and industry efforts, this market lacks a proper method to estimate returns. Consequently, and not surprisingly, a universally accepted index to monitor prices in the market for paintings has yet to emerge. In this article, we propose a metric that we think it can fulfill this need.

In what follows we provide some context for the challenges associated with estimating returns in the case of paintings. We also review very briefly the issues related to the two main approaches, namely, repeat-sales regression models (RSRM) and hedonic pricing models (HPM). Then we introduce a new metric to estimate returns, we provide a mathematical justification for it, and we show that it satisfies the monotonicity condition (a key requirement to build a credible price index). Finally, we show with a simple example the benefits of the new metric, and also, the problems associated with estimating returns and building price indexes with the time-dummies of conventional the HPMs.

# BACKGROUND

The main challenge to estimate returns in the case of paintings—and subsequently, to construct a price index— is that there is no clear definition of what constitutes the *real* (actual or true) return. There is, of course, a trivial case: if we are dealing with only one painting that has been sold twice, the computation (and lack of ambiguity) is obvious. If, on the other hand, we are dealing with a group of price-observations referring to several paintings, the matter gets messy very quickly. We need to deal with product variety and the fact that for many paintings there will be just one observation. In summary, unlike the case of stocks, commodities, and bonds—where there is a unique and well-defined price—the market for paintings lacks this basic building block. Previous researchers have attempted to tackle this challenge using two approaches: RSRMs and HPMs.

Conceptually, the RSRM approach is very clean. The idea is simply to compute the return associated with pairs of "matched-sales" (paintings with two price-observations at different times) and then combine these returns to obtain some value representative of all the data considered. Some researchers think, erroneously, that this approach controls for paintings characteristics. Actually, it does not for the returns computed are total returns: they capture price variations resulting from both, the features of the paintings considered as well as market dynamics. In fact, this is precisely the strength of this method, which, incidentally, has been successfully used in the U.S. real estate market. Unfortunately, in the case of paintings, this approach has two drawbacks. Unlike the U.S. real estate market, the number of repeat-sales observations is a very small set compared to the universe of all sales. In addition, the repeat-sales set suffers from a serious selection bias problem, i.e., it is not a random sample of the universe. Thus, this technique, despite its sound theoretical basis, is of little practical use in the market for paintings. These issues have been recognized by most if not all researchers in the field.

A second approach, perhaps the most popular in the art market, is to rely on HPMs. The idea, again, is sound and simple: it consists of expressing the logarithm of the price of a painting in terms of a polynomial expansion based on discrete or continuous quantitative variables associated with the attributes of each painting plus a set of time dummies to reflect the year in which the painting was sold. Typically, a HPM is constructed using all the available data for the time-periods considered, and the returns and associated values of the price index are estimated based on the time-dummy coefficients of the regression. From the operational viewpoint, this approach has some problems. The variability explained (adjusted $R^2$) by these models, especially in the case of individual painters, is often low. And, the estimated coefficients of these HPMs are prone to show instabilities.

However, it is conceptually that this approach presents two very severe —and unsurmountable— shortcomings. The first issue has to do with the interpretation of the returns obtained with this approach. Clearly, unlike the case of RSRMs, these are not total returns for one has controlled for the paintings characteristics. These are returns associated with an *ideal* (average) painting whose attributes remain invariant over time. The practical significance or interpretation of a return associated with such *ideal* painting is still a matter of debate. The second issue, far more serious than the first, is that returns computed with this method lead to price indexes that do not satisfy the monotonicity axiom (Fischer, 1922). This is a *sine qua non* condition for any price index. Simply stated, this implies that price increases at the product level (keeping the product reference set constant) cannot result in a decrease in the value of the index. The awkward violation of this very important condition is a problem that mars all hedonic time-dummies based-indexes, not just art-market indexes. However, even though this problem has been known for at least ten years, art-related researchers have conveniently ignored the issue.

Clearly, there is a need to have a better way to compute returns and price indexes for this market. We hope the metric we introduce in the next section can be of some help. Finally, Ginsburgh, Mei, and Moses, (2006) provide an excellent review of most issues related to HPMs, RSRMs, and price indexes in general, plus an extensive list of references. Another useful reference is the European Commission (2013) handbook on consumer price indexes. In this brief overview, we have purposely avoided taxing the reader with a tedious list of references at the end of every sentence.

## THE METRIC AND THE INDEX

### Intuition behind the Metric and the Index

Paintings, notwithstanding their artistic qualities, are essentially two-dimensional objects. Based on this consideration, it makes sense to express the value of a painting not using its price but rather its price per unit of area (in this study, US dollars per square centimeter). By normalizing the price, this metric intends to offer the investor a financial yardstick that goes beyond the price, while not attempting to control for the specifics of the painting beyond its area.

The intuitive appeal of this metric is obvious: simplicity, ease of computation, transparency, and straightforwardness. In fact, there is already a well-established precedent for this approach. For example, prices of other two-dimensional assets, such as raw land, are frequently quoted this way (e.g. dollars per acre, or euros per hectare). The same approach is sometimes used to quote prices of antique rugs.

More formally, let $P$ be the observed price of a painting at some specific time and $A$ its area. We define $p$, its unitary (or normalized) price as $p=P/A$. If we have a set of $N$ paintings, we claim that a representative price, $p^*$, for such set can be computed as

$$p^* = \left(\prod_{i=1}^{N} p_i\right)^{1/N} \qquad (1)$$

in which $p^*$ obviously is the geometric mean of the corresponding $p_i$'s. It follows then, that the price variation between two consecutive periods, 0 and 1, can be expressed as $p^*_1/p^*_0$. This is the basis for creating a price index. We can set the value of the index at whatever arbitrary value we wish initially and then build the sequence based on the period-by-period variation of the $p^*$'s ratios. The triviality of this computation is apparent. The solid foundation on which it is based —in the case of paintings— is not so obvious.

**Mathematical Foundation**

We start with the conventional HPM as it applies to paintings. Specifically, we write

$$Log P_n^t = \alpha + \sum_k \beta_K z_{nk} + \sum_i \delta^i \tau_n^i + \epsilon_n^t \quad (2)$$

where $P_n^t$ is the price of painting $n$ at time $t$; $\alpha$ is a constant; $\beta_k$ represents the painting characteristics parameters; $z_{nk}$ denotes the value of characteristic $k$ in the case of painting $n$; $\tau_n^i$ is a time dummy (1 if painting $n$ is sold at time $t$, 0 otherwise); and $\varepsilon_n^t$ is the error term. The time dummy parameters $\delta^i$'s are often used to build a price index as the ratio $exp(\delta^a)/exp(\delta^b)$ captures the price variation between periods $a$ and $b$. This is the usual HPM formulation in the context of paintings.

Suppose now that we rely on only one characteristic—the logarithm of the area—and we force the corresponding $\beta$ to be 1. Then, equation (2) turns into

$$Log P_n^t = \alpha + (1)Log(A_n) + \delta^t + \epsilon_n^t \quad (3)$$

which, after some algebra leads to

$$Log\, p_n^t = Log(\frac{P_n^t}{A_n}) = \alpha + \delta^t + \epsilon_n^t \quad (4)$$

In essence, estimating returns (and price indexes) by relying on unitary (normalized) prices can be interpreted as using a special version of a hedonic model. That is, a model (equation (4)), in which no characteristics are used on the right-hand side (the area is implicitly participating, via the unitary price, on the left-hand side of the equation). Consequently, the returns estimated with the metric proposed, can be interpreted as returns estimated with the $\delta^i$'s of this particular hedonic model. In this sense, the metric proposed falls within the standard framework of the conventional HPM approach, with one caveat: the simplicity of the computation due to the particular structure of the HPM.

Recall also, that when OLS are employed, the usual HPM time-dummy price index variation between two periods, 1 and 0, can be expressed as follows (in reference to equation (2))

$$Index\ (1,0) = \left[\frac{\left(\prod_{i=1}^{N(1)} P_i^1\right)^{1/N(1)}}{\left(\prod_{i=1}^{N(0)} P_i^0\right)^{1/N(0)}}\right] \theta \quad (5)$$

where *N(1)* and *N(0)* denote the number of observations in each period and $\theta$ is defined as

$$\theta = exp\left[\sum_k \beta_K (\varphi_k^0 - \varphi_k^1)\right] \quad (6)$$

with $\varphi_k^0$ and $\varphi_k^1$ representing, for paintings in either period, 1 or 0, the average (arithmetic mean) value of the corresponding characteristic ($z_{ik}$) where i=1, …, *N(0)* or *N(1)* depending on the case.

Equation (5) simply states that the price index variation is the product of (i) the ratio of the geometric mean of the prices; and (ii) a factor ($\theta$) that adjusts for changes in the value of the characteristics. Obviously, $\theta$=1, means no adjustment. Therefore, the price index proposed ($p^*_1/p^*_0$) can be interpreted, by invoking equation (4), as a regular HPM-based time-dummy index, which is constructed with a very special HPM. The fact that $\theta$=1 (no correction for characteristics in our proposal) is consistent with the fact that the correction already took place by virtue of computing the ratio based on unitary prices (that is, after dividing the prices by the area). In other words, the metric we propose and the corresponding price index computation, can be regarded just as a special case of equation (5).

**Monotonicy Condition**

The monotonicity condition can be stated as follows. Suppose **X** is a vector of prices associated with a given set of observations. Let **λ** be a vector whose entries are either 0 or positive. A price index function, $\rho$, satisfies the condition if $\rho(\mathbf{X}+ \boldsymbol{\lambda}) \geq \rho(\mathbf{X})$. That is, if the prices of one or several items increase (keeping the characteristics unchanged) the price index has to remain the same or increase, but cannot decrease.

The price index proposed clearly satisfies the monotonicy condition as it is apparent from equation (1) that increasing the value of any $p_i$ results in a positive increment in $p^*$.

In summary, we have proposed a price index for the art market of paintings that is intuitive, easy to compute, free of computational ambiguities, and stable. Furthermore, we have also shown that this index is very similar —in spirit and in computation— to a particular type of hedonic model. That is, it rests on a sound theoretical foundation. However, unlike the price indexes based on the time-dummy coefficients of conventional hedonic models, it does not violate the monotonicity condition. We state that these attributes makes this price index an attractive contender to the more conventional approaches.

# EXAMPLE OF APPLICATION

We consider two sets of observations (A and B), which are described in Table 1. The data reflect realized Renoir's paintings auction prices (and their corresponding characteristics) observed during the 1989-1990 period. The information was taken from the artnet database and the prices are premium prices expressed in 2010 US dollars. The datasets were constructed to be different in terms of the area (lower values in dataset A). Dataset C (not included in Table 1) is identical to dataset B except that we have purposely increased in 50% the sale price of the 29[th] observation. For convenience, we set the price level associated with dataset A at 100.

**Table 1. Example Datasets: from Pierre-Auguste Renoir 1989-1990 auction sales**

| Obs. Number | Dataset | P Price (US$) | A Area (cm$^2$) | Height/Width Ratio | p* Price/Area (US$/cm$^2$) |
|---|---|---|---|---|---|
| 1  | A | 105,771     | 74.90    | 1.129 | 1,412.16  |
| 2  | A | 107,809     | 187.00   | 0.647 | 576.52    |
| 3  | A | 132,526     | 202.16   | 1.043 | 655.55    |
| 4  | A | 141,992     | 313.50   | 1.152 | 452.93    |
| 5  | A | 294,025     | 847.00   | 0.691 | 347.14    |
| 6  | A | 396,037     | 658.56   | 0.583 | 601.37    |
| 7  | A | 404,320     | 193.04   | 1.197 | 2,094.49  |
| 8  | A | 609,276     | 659.34   | 0.595 | 924.07    |
| 9  | A | 645,281     | 494.00   | 1.368 | 1,306.24  |
| 10 | A | 738,129     | 2,576.00 | 0.821 | 286.54    |
| 11 | A | 738,354     | 535.60   | 1.262 | 1,378.55  |
| 12 | A | 823,315     | 645.00   | 1.395 | 1,276.46  |
| 13 | A | 1,325,251   | 749.30   | 1.161 | 1,768.65  |
| 14 | A | 2,295,088   | 942.50   | 1.121 | 2,435.11  |
| 15 | B | 1,401,186   | 3,905.00 | 1.291 | 358.82    |
| 16 | B | 1,666,029   | 1,346.38 | 0.789 | 1,237.41  |
| 17 | B | 1,704,025   | 2,882.00 | 1.489 | 591.26    |
| 18 | B | 2,048,648   | 2,576.00 | 0.821 | 795.28    |
| 19 | B | 2,213,831   | 1,353.00 | 1.242 | 1,636.24  |
| 20 | B | 2,419,564   | 3,752.00 | 0.836 | 644.87    |
| 21 | B | 3,252,946   | 756.00   | 1.313 | 4,302.84  |
| 22 | B | 3,326,901   | 3,515.40 | 1.206 | 946.38    |
| 23 | B | 3,658,125   | 1,312.00 | 1.281 | 2,788.21  |
| 24 | B | 4,453,582   | 2,540.92 | 1.217 | 1,752.74  |
| 25 | B | 4,453,582   | 8,804.25 | 0.389 | 505.84    |
| 26 | B | 9,224,292   | 3,597.00 | 1.211 | 2,564.44  |
| 27 | B | 11,401,170  | 3,499.20 | 1.200 | 3,258.22  |
| 28 | B | 29,393,641  | 8,100.00 | 1.235 | 3,628.84  |
| 29 | B | 146,841,502 | 8,892.00 | 0.684 | 16,513.89 |

The idea is to estimate prices indexes $I_{BA}$ and $I_{CA}$ for datasets B and C (using A as reference) with: (i) the approach suggested in this article; and (ii) the time-dummy coefficients method based on a suitable HPM. Return and price index computations based on our metric are straightforward. In order to estimate comparable figures using the time-dummies approach we rely on the HPMs described in Tables 2 and 3. They employ two characteristics: the area of the painting and its height/width ratio. The models are satisfactory as they show a reasonable explicatory power ($R^2$'s higher than 70% in both cases).

**Table 2.  Results from the HPM with datasets A and B**

|                    | Coefficients | Standard Error | t Stat | P-value   |
|--------------------|--------------|----------------|--------|-----------|
| Intercept          | 11.619049    | 0.7046         | 16.49  | 6.053E-15 |
| Area ($cm^2$)      | 0.000411     | 9.3484E-05     | 4.39   | 0.00018   |
| Height/Width Ratio | 1.051534     | 0.6297         | 1.67   | 0.10740   |
| Time Dummy         | 1.068575     | 0.4522         | 2.36   | 0.02622   |

**Table 3.  Results from the HPM with datasets A and C**

|                    | Coefficients | Standard Error | t Stat | P-value   |
|--------------------|--------------|----------------|--------|-----------|
| Intercept          | 11.624505    | 0.7321         | 15.88  | 1.44E-14  |
| Area ($cm^2$)      | 0.000429     | 9.71E-05       | 4.42   | 0.00017   |
| Height/Width Ratio | 1.034313     | 0.6543         | 1.58   | 0.12647   |
| Time Dummy         | 1.038821     | 0.4699         | 2.21   | 0.03642   |

Figures 1 and 2 display the results and they are troubling. The index associated with our metric satisfies the monotonicity condition, namely, $I_{CA} > I_{BA}$; the HPM time-dummies based-index does not satisfy it. Just to be clear: although the prices associated with dataset C are higher than those of dataset B, the time-dummies based-index, $I_{CA}$, is lower than $I_{BA}$.

**Figure 1.** Price Index computation based on the geometric mean of the normalized prices, for two different datasets (B and C), and using the same reference dataset (A). Prices for datasets B and C are identical except that $P^C_{29} = (1.5) \times P^B_{29}$.

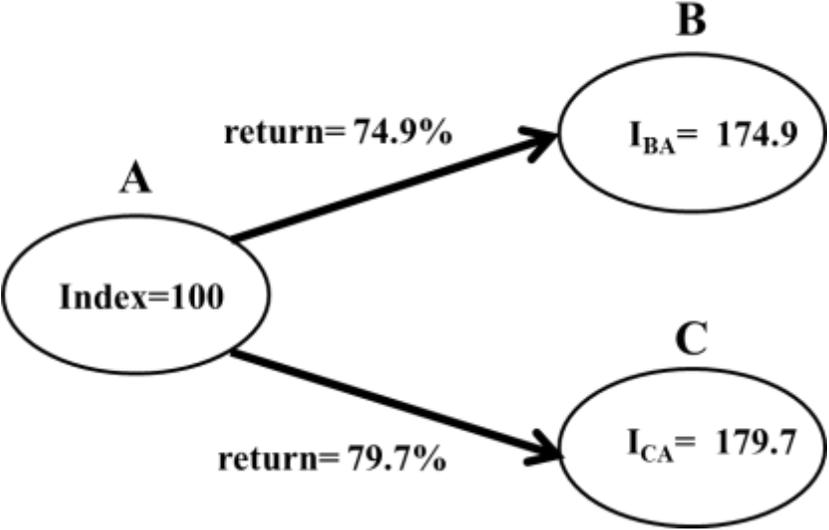

**Figure 2.** Price Index computation based on the time dummies of the hedonic model, for two different datasets (B and C), and using the same reference dataset (A). Prices for datasets B and C are identical except that $P^C_{29} = (1.5) \times P^B_{29}$.

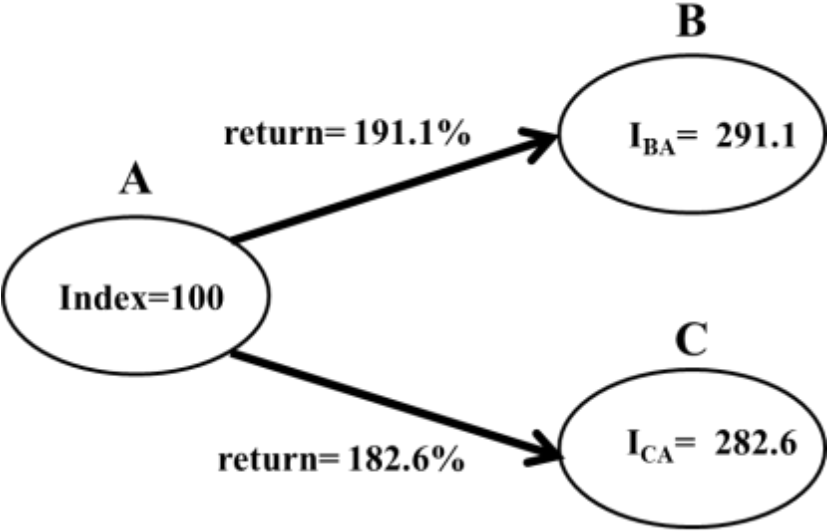

It might be tempting to believe that this undesirable behavior is entirely an artifact of this example. But, this example just highlights an unpleasant fact that has been known for quite some time and is intrinsic to the time-dummies method (Melser, 2005). In fact, increasing the price of observations 25 or 28 (instead of 29), while holding the remaining prices unchanged, we can also achieve the same effect: obtaining an overall price increase that, in the case of the time-dummies, results in an index whose value is smaller than $I_{BA}$.

The need for a price index to meet the monotonicity condition, of course, is not a novel idea, it goes back to almost a century ago and it was well established by Fisher in his classical book The Making of Index Numbers (Fisher, 1922). The rationale for this requirement, leaving aside the mathematical complexities, is obvious: a price index that can move in the opposite direction of price levels is not only counterintuitive and difficult to explain. It is simply useless and misleading.

Melser (2005) has pointed out that a precondition to violate the monotonicity requirement is the existence of a relationship between the time-dummies (in our example, dataset dummies) and one of the characteristics employed in the HPM. Figure 3 illustrates this point: there is indeed a significant relationship between the dataset variable and paintings' area (*A*) in this example. We can speculate ─and this is a conjecture─ that in the case of more conventional products such as computers or other consumer goods, the possibility of a strong link between the time-dummies and the characteristics might be small. Unfortunately, in the art market, where paintings characteristics can exhibit significant changes from period-to-period, the likelihood of encountering monotonicy-condition violations is probably higher. This fact certainly undermines the potential validity of previous studies based on such indexes.

**Figure 3.** Relationship between the paintings area and datasets A and B.

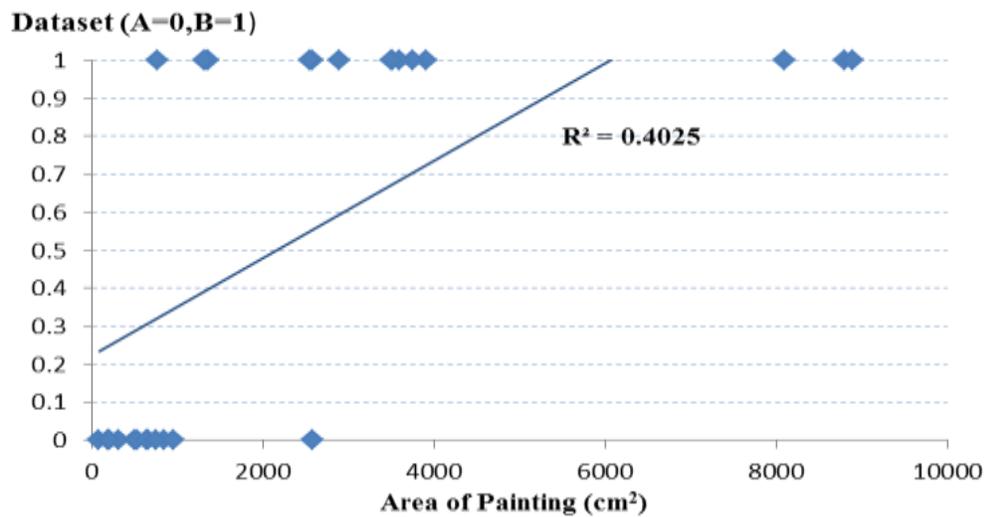

## CONCLUSIONS

Two conclusions emerge from this study.

[1] The metric introduced herein, and the resulting price index, have several appealing features: they are intuitive, easy to compute, and do not exhibit ambiguities or lack of stability problems. In addition, they rest on a solid foundation as they can be interpreted as a particular type of hedonic model which −and this is critical− meet the monotonicity condition.

[2] The example highlighting that price indexes based on the time dummies of HPMs violate the monotonicity condition is quite revealing. This is not a peculiarity of the art market; it is a mathematical property intrinsic to the HPM time-dummies-based indexes. We certainly do not claim to have discovered this fact. Previous researchers have acknowledged this problem, albeit in different applications, but not in the context of paintings. However, in the market for paintings, this violation is particularly unsettling, since a substantial part of current and past academic research is based on price indexes derived from HPMs. What are we supposed to make of these findings? Were they based on actual market movements or artifacts of these models? It is an open question.

Moreover, let us consider that:

- Time-dummies based-indexes (notwithstanding the monotonicity issue) reflect the price variations of an *ideal* painting whose properties remain unchanged over time, and, as a result, they are difficult to interpret; and

- Behind the conventional HPM formulation (see equation (2)), there are two seldom-recognized assumptions, both of which, are strongly violated in the art market. Namely, the fact that the utility function implicit in the formulation is time invariant, and, is the same for all art investors or collectors. Clearly, in reference to more conventional consumer goods one could argue that these two violations are more likely to be weak. However, in the art market, where supply and demand as well as trends experience wild swings from year to year and tastes differ greatly from one collector or investor to another, these assumptions are obviously unrealistic. The argument that one could use a sequence of adjacent HPMs (instead of one global model for the entire dataset) to at least capture the period-to-period variation in tastes (utility function) is not very practical. The reason is that scarcity of data often prevents this exercise. And even if this were the case, the very fact that the periods typically used in art market analyses are long −years instead of weeks or months− contributes to undermine the assumption of time-invariant tastes.

In summary, in the context of paintings, the credibility of HPM time dummies-based indexes appears to suffer from too many weaknesses to be taken seriously. One cannot help but wonder if they are worth the trouble. In fact, it might be better to just abandon them and move on.

# REFERENCES

European Commission, EUROSTAT. (2013). Handbook of Residential Property Indices.

Fischer, R. (1922). The Making of Index Numbers; a Study of their Varieties, Tests, and Reliability. Boston: Houghton-Mifflin.

Ginsburgh, V., Mei, J. and Moses, M. (2006). The Computation of Price Indices, Chapter 27, in Handbook of the Economics of Art and Culture, Volume I, Elsevier, B.V.

Melser, D. (2005). The Hedonic Regression Time-Dummy Method and the Monotonicity Condition. Journal of Business & Economic Statistics, Volume 23, Number 4, pp. 485–492